\documentstyle[aps,feyn,epsf]{revtex}

     \newcommand{\be}{\begin{equation}}
     \newcommand{\ee}{\end{equation}}
     \newcommand{\bea}{\begin{eqnarray}}
     \newcommand{\eea}{\end{eqnarray}}

\begin{document}
\title{ A look at the crossover region between BCS superconductivity and
    Bose Einstein Condensation}
\author{ P. A. Sreeram and Suresh G. Mishra}
\address{Institute of Physics, Sachivalaya Marg,
     Bhubaneswar 751 005, India.}

\maketitle

\begin{abstract}
    Pair fluctuation theory has been used to study the crossover
    from the weak coupling BCS theory to the strong coupling  
    Bose Einstein Condensation. The effect of fluctuations has been
    studied over the whole crossover regime. It has been shown that the 
    pair fluctuations are enhanced considerably in both two and three
    dimensions and hence mean field theory is inadequate to study the
    physical properties in this regime. A self consistent scheme for
    calculating the pair susceptibility is given.
\end{abstract}
\vspace{1cm}

    The study of crossover from BCS to Bose Einstein Condensation has
    been a subject of intense study in recent times. The motivation
    for this study has been from the point of view of trying to model
    High $T_c$ superconductors. These materials share many properties
    with BCS superconductors, however, the coherence length $\xi$ in
    these materials is unusually small \cite{Ausloos97}. The short
    coherence length feature is also apparent in the phenomenon of Bose
    Einstein Condensation (BEC) where the coherence length $\xi$ is
    expected to be of the order of an atomic spacing. Moreover, in the 
    BEC process the temperature at which pairing occurs is different
    from the pair condensation temperature. The phenomenon of BEC was
    was known as Schafroth condensation \cite{Blatt64} in the pre BCS
    days of superconductivity. Recently this has lead to a formulation
    of theories based on occurance of pseudo gaps in the normal state
    of these materials. \cite{Timusk99,Emery97,Pines97,Noz99}The
    crossover from BCS to BEC has been studied 
    as a case of weak to strong attractive coupling theory in details
    during recent years within the mean field theory or RPA
    \cite{Mohit95,Eagles69,Leggett80,Nozieres85,Pistolesi94,Haussmann93,Stinzing97}.   
    These results agree with the expected behavior at the two extreme
    limits (i.e. the weak coupling BCS limit and the strong coupling BEC
    limit) and predict a smooth crossover from one limit to the
    other. However, the intermediate regime, to which the High $T_c$
    superconductors are believed to belong within this theory, has not
    been well understood. 
    The theory is based on the existence of two parameters which can be
    varied - the electronic density $n$ and the coupling strength
    $g$. Coupled self consistent equations for these two parameters leads
    to the calculation of $T_c$, the transition temperature and the
    chemical potential $\mu$. For a constant density, if the coupling
    strength is varied, there is  smooth crossover from the weak coupling
    BCS superconductivity to the strong coupling Bose Einstein
    condensation. In principle, however, both the parameters can be
    varied. If this is done such that $T_c$ remains constant and $\mu$ is
    varied there is a crossover with respect to $\mu$, where for large
    positive $\mu$ ($\mu \sim \epsilon_F$, the Fermi temperature) the
    results for the BCS superconductors are recovered and for large
    negative $\mu$ the BEC results are recovered. Thus, the intermediate
    regime can be interpreted in this regime as the region close to $\mu =
    0$. In what follows, we shall take this point of view in order to
    study the crossover.

    In this work, we wish to study the crossover regime from the point of
    view of pair fluctuations. The main aim of this work is to ask the
    question - how important are fluctuation contributions to this
    phenomenon of crossover, especially in the region close to the
    crossover ? As has already been mentioned before, most of the studies
    in this subject have taken recourse to mean field theory. If
    fluctuation contributions do become important, then, mean field theory 
    is bound to breakdown. We therefore set up an equation for the
    pair susceptibility within the mean fluctuation field
    approximation (MFFA) where the effect of fluctuations are included
    in a self consistent manner. This self consistent equation is then
    solved to obtain the inverse pair susceptibility over the whole
    regime of crossover from the BCS limit (where $\mu \gg T$) to the
    BEC limit (where $-\mu \gg T$).   

    The starting point for the calculation is the BCS Hamiltonian in the
    momentum space with an s-wave attractive potential which is given by,
\begin{equation}
\label{eq:Hubb-2}
    H-\mu N = \sum_{p,\sigma}\epsilon_p C^{\dagger}_{p,\sigma}
    C_{p,\sigma} - \frac{1}{2} g \sum_{p,p^\prime,q,\sigma} 
    C^{\dagger}_{p+q,\sigma}C^\dagger_{-p,-\sigma}
    C_{-p^\prime,-\sigma}C_{p^\prime+q,\sigma},
\end{equation}
    where $\epsilon_p = p^2/2m$ is the free electron kinetic energy,
    and $g $ is the interaction strength. The mean field results for
    this Hamiltonian are well known for both the BCS limit and the BEC 
    limit.  

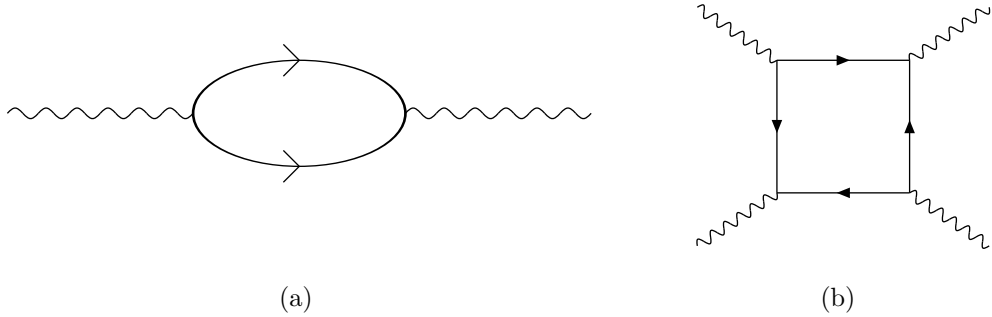
\begin{figure}
   
    \begin{picture}(300,100)(200,100)
    \Photon(210,150)(280,150)2 6
    \Text(320,80)[c]{(a)}
    \Oval(320,150)(20,40)(0)
    \Line(314,176)(320,170)
    \Line(314,164)(320,170)
    \Line(314,124)(320,130)
    \Line(314,136)(320,130)
    \Photon(360,150)(430,150)2 6
     \ArrowLine(500,170)(550,170)
    \ArrowLine(500,170)(500,120)
    \ArrowLine(550,120)(500,120)
    \ArrowLine(550,120)(550,170)
    \Photon(470,190)(500,170)2 6
    \Photon(470,100)(500,120)2 6
    \Photon(580,190)(550,170)2 6
    \Photon(580,100)(550,120)2 6
    \Text(525,80)[c]{(b)}
    \end{picture}

    \vspace{0.9cm}
    \caption{Lowest order diagrams for Free Energy}
    \label{fig:k2}
    \end{figure}

     The functional integration technique is a convenient method for
     treating the fluctuation problem microscopically
     \cite{Mishra78}. For the case of superconductors the technique
     was developed by Langer \cite{Langer64}, Rice \cite{Rice67} and
     others. We follow the procedure adopted by Rice in what
     follows. Using the functional integral technique the partition
     function can be expressed as, 
    \begin{equation}
    \label{eq:part-func-3}
    Z/Z_0  = \int \prod_{q,m} \frac{{\cal D} x_{q,m}}{\pi} \exp
    \left \{ -\beta F[x_{q,m}]\right \} 
    \end{equation}
    where $Z_0 $ is the partition function of non-interacting system
    and $x_{q,m}$ is the (bosonic) auxiliary field with momentum $q$
    and frequency $\omega_m$. The pairing correlation is the only one
    which has been considered here. A diagrammatic expansion of $ F [
    x_{q,m} ]$ in powers of $ x_{q,m} $ is possible and
    leads to the Ginzburg Landau functional. The corresponding
    expression for the free energy functional is given by,   
    \begin{equation}
    \label{eq:y1}
    F^{(1)}[x_{q,m}] = \frac {1}{\beta} \sum_{q,m} (1 + g
    K^{(1)}_q)x_{q,m}^\star x_{q,m} + 
    \frac {1}{\beta} g^2 \sum_{q,m} K^{(2)}_{q_1,q_2,q_3}
    x^\star_{q_1} x^\star_{q_2} x_{q_3} x_{q_1 + q_2 - q_3} .
    \end{equation}
    These terms are shown in Fig. (\ref{fig:k2}).
    The Eq. {\ref{eq:y1}} represents the free energy of interacting
    electrons near a superconductive instability in terms of
    interacting pair fluctuation field. The first term is the ``free'' 
    or the RPA term (Fig. \ref{fig:k2}a), while the second term
    represents the interaction between pair fields
    (Fig. \ref{fig:k2}b). Parameters of this model, e.g. transition
    temperature, collective mode dispersion, fluctuation spectrum,
    fluctuation coupling vertices are determined by properties of the
    underlying fermion system. Expanding $K^{(1)} $ and $ K^{(2)} $ in
    powers of $q$ and $\omega_m$ and retaining the lowest powers we get,
    \begin{equation}
    \label{eq:y-3}
       \beta F[x_{q,m}] = \left [ a + c q^2 - \imath\omega_m
       \right ] \mid x_{q,m} \mid ^2 + b \mid
       x_{q,m} \mid^4 . 
     \end{equation}
    The coefficients $a$, $b$, $c$ and $d$ are given by, 
    \begin{eqnarray}
    \label{eq:a,c,d-1}
    a & = & \frac{1}{g} - \frac{1}{\beta}\sum_{p,\nu_l}\frac{1}
    {(( \epsilon_p - \mu )^2+\nu_l^2)} \nonumber \\
       b & = & \frac{1}{2\beta} \sum_{p\sigma,l}
       \frac{1}{\left [ (\epsilon_p - \mu )^2 - \nu_l^2 \right
          ]^2} \nonumber \\ 
     c & = &
    \frac{2m}{\beta}\sum_{p,\omega_l}\frac{1}{( (\epsilon_p -
          \mu)^2+\nu_l^2)^2}  
    \Biggl[ (\epsilon_p - \mu ) - \eta \epsilon_p \frac{((\epsilon_p - 
      \mu )^2 -\nu_l^2)} 
    {((\epsilon_p - \mu)^2+\nu_l^2)}\Biggr]  \nonumber \\      
    d & = & \frac{1}{\beta}\sum_{p,\nu_l} \frac{(\epsilon_p - \mu 
      )}
    {((\epsilon_p -\mu )^2+\nu_l^2)^2},
    \end{eqnarray}
    where $\nu_l = (2l+1)\pi/\beta$. The value of $ \eta $ comes out
    to be 2 in 2D and $4/3 $ in 3D. At $T = T_c$, $a = 0$,  which
    leads to the Thouless criterion, $ ( 1 + g K^1_0 = 0) $,
    \begin{equation}
    \label{eq:1bygfroma}
    \frac{1}{g} = \frac{1}{\beta_c}\sum_{p,\nu_l}\frac{1}
    {((\epsilon_p - \mu_v)^2+\nu_l^2)} = \frac{1}{2} 
    \int_0^\infty d\epsilon ~ \rho (\epsilon) 
      \frac{\tanh[\beta_c(\epsilon-\mu_c)/2]}{(\epsilon - \mu_c)}
    \end{equation}
    The integral on the right hand side is UV divergent. In the
    conventional weak coupling theory this is taken care of by putting
    an upper cutoff at the Debye energy. However, since we are
    interested in the crossover from weak to strong coupling behavior
    this restriction on the energy integration has to be removed. The
    singular behavior is taken care of by replacing the bare $ g $ by
    the low energy limit of the two body $ T $ matrix. 
    \begin{equation}
    \label{eq:1tmatrix}
    \frac {1}{g} = - \frac {1}{t} + \sum^{\prime} \frac {1}{2
      \epsilon_k },
    \end{equation}
    where $ 1/t $ turns out to be $ m/4 \pi a_s $ for $ d = 3 $ and $
    (m/4 \pi ) \ln ( E_b /2 ) $ for $ d = 2 $. Here $ a_s $ is the
    s-wave scattering length and $ E_b = 1/ma_s^2 $. 

    The number density of electrons $n$ and the coupling strength $g $  
    determine the transition temperature and also specify whether one
    is in Bose or the BCS regime of pairing. These are related to the
    transition temperature $T_c$ and the chemical potential $\mu$
    through Eqs. \ref{eq:1bygfroma}, {\ref{eq:1tmatrix} and the
    equation for the number density, viz.  
    \begin{equation}
    \label{eq:1number}
    n = \sum_k ( 1 - \tanh \frac {\beta ( \epsilon_k - \mu )}{2} ),
    \end{equation}
    which are to be solved self consistently. One can consider ($n$
    and $g$) or, ($T_c$ and $\mu$) combination as the parameters in
    the problem. We find convenient to consider the latter pair,
    particularly when dealing with the crossover region. The region
    around $\mu \sim 0$ marks the crossover. We write $a$ etc. in terms
    of $ T_c $ and $ \mu_c$ , i.e.,      
    \begin{equation}
    a = \frac{1}{2}\left(\int_0^\infty d\epsilon \rho(\epsilon)
         \frac{\tanh[\beta_c(\epsilon-\mu_c)/2]}{(\epsilon - \mu_c)} -
         \int_0^\infty d\epsilon \rho (\epsilon)
         \frac{\tanh[\beta(\epsilon-\mu)/2]}{(\epsilon - \mu)} \right).
    \end{equation}
    The expressions for $b, c$ and $d$ are nonsingular  at $T_C$.
  
    The pair susceptibility can be calculated in this scheme as, 
     \begin{equation}
       \label{eq:pair-susc-0}
       \chi(q,\omega_m) = <\mid x_{q,m} \mid^2> = \int dx_{q,m}
       dx_{q,m}^\star \mid x_{q,m} \mid^2 \exp{-\beta F[x_{q,m}]} /
       \int dx_{q,m} dx_{q,m}^\star exp{-\beta F[x_{q,m}]}.
     \end{equation}
     We now write down a self consistent equation for the inverse of
     pair susceptibility within the mean fluctuation field
     approximation. This amounts to writing the quartic term, in  
     the expansion of the free energy functional in terms of the pair
     field, as a quadratic form (i.e. $\mid x_{q,m}\mid^4 = <\mid
     x_{q,m}\mid^2> \mid x_{q,m} \mid^2$) and taking average over the
     effective Gaussian distribution. The technique is well known
     \cite{Chaikin97}; we, however, follow the earlier work on spin
     fluctuations \cite{Mishra78}. The pair susceptibility is given by,     
     \begin{equation}
      \label{eq:pair-susc}
      \chi(q,\omega_m) = \frac{1}{\alpha(T, \mu) + c q^2 + i d \omega_m},
    \end{equation}
     where,
     \begin{equation}
     \label{eq:alpha-bare}
     \alpha(T,\mu) = a + b \sum_{q^\prime,\omega_{m^\prime}}
     \chi(q^\prime,\omega_{m^\prime}).
     \end{equation}
     
    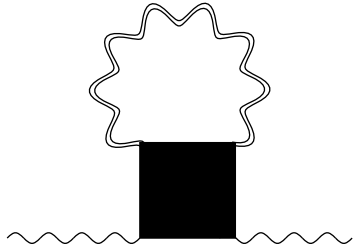
\begin{figure}
    \begin{picture}(300,110)(165,30)
    \Photon(330,49)(380,49)2 4
    \GBox(380,49)(416,85)0
    \Photon(410,49)(460,49)2 4
    \PhotonArc(395,105)(28,-45,244)4 8
    \PhotonArc(395,105)(30,-45,244)4 8
    \end{picture}
    \caption{Mean fluctuation field diagram}
    \label{fig:mffa}
    \end{figure}

    To calculate the susceptibility one needs values of
    Ginzburg-Landau coefficients, $a$, $b$, $c$ and $d$. On 
    performing the summation over the Matsubara frequencies, one gets
    in 3 dimensions, 
    \begin{equation}
\label{eq:3d} 
  b = \frac{1}{8} \int_0^\infty d\epsilon ~ \rho_{3D}(\epsilon)
      \Phi(\epsilon), ~~~
     c  =   \frac{1}{12} \int_0^\infty d\epsilon ~ \rho_{3D}(\epsilon) 
 \epsilon ~ \Phi(\epsilon), ~~~\rm{and}~~~ 
      d = \frac{1}{8} \int_0^\infty d\epsilon ~ \rho_{3D}(\epsilon) 
      (\epsilon - \mu) ~ \Phi(\epsilon). 
    \end{equation} 
     Here, $\rho_{3D}(\epsilon) = (2m)^{3/2}\sqrt{\epsilon}/4\pi^2$ is 
     the density of states in 3 dimensions and 
     \begin{equation}
 \Phi(\epsilon) =  \{ 2 \tanh (\beta (\epsilon-\mu) /2) - 
       \beta (\epsilon - \mu) {\rm sech}^2(\beta_c (\epsilon -
       \mu )/2) \} /(\epsilon - \mu)^3.
\end{equation}
     It is not possible to evaluate the energy integral analytically
     throughout the parameter space. Only in the extreme limits the
     results are available. In the BCS limit  $\beta \mu \gg
     1$,
     $ a = \ln(\pi e^2/ 8e^\gamma \beta \mu) \rho_{3D}(\mu )) -
     \ln(\pi e^2 /8e^\gamma \beta \mu_c ) \rho_{3D}(\mu_c)  $,  
     $ b = (7\zeta(3) /8\pi^2 ) \rho_{3D}(\mu) \beta^2 $
     $ c = (7\zeta(3)/12 \pi^2 )\rho_{3D}(\mu ) \mu  \beta^2 $, and 
     $ d = \ln  (8e^\gamma \beta \mu/\pi) \rho_{3D}(\mu )/4\mu $ .
     In the Bose Einstein regime ($- \beta \mu \gg 1$) the
     corresponding value are, 
     $ a = ( \pi /2 ) ( \rho_{3D}(\mid \mu \mid)- \rho_{3D}(\mid \mu_c
     \mid ))$,       
     $ b =  \rho_{3D}(\mid \mu \mid)( \pi / 32 \mid \mu \mid ^2 ) $,     
     $ c = \rho_{3D}(\mid \mu \mid )( \pi / 16 \mid \mu \mid ) $, and  
     $   d = \rho_{3D}(\mid \mu \mid) ( \pi   / 8  \mid \mu \mid ) $.

     Similarly, in two dimensions, the summation over the Matsubara
     frequencies gives,
   \begin{equation}
   \label{eq:2d}
    b  =  \frac{1}{4}\int_0^\infty d\epsilon ~
     \rho_{2D}(\epsilon) \Phi(\epsilon),~~~
       c   =  \frac{1}{4} \left(\frac{7\zeta(3)}{2\pi^2}\beta^2 \mu
      \theta(\mu) + \mid \mu \mid \int_{\mid \mu \mid}^\infty d \xi
      \rho_{2D}(\epsilon)
      \frac{\tanh(\beta\xi/2)}{\xi^3}\right),~~~\rm{and}~~~
     d =   \rho_{2D}(\epsilon) \frac{\tanh(\beta\mu/2)}{4 \mu},
\end{equation}
     where $\rho_{2D}(\epsilon) = m/2\pi$. Again, in the BCS limit,  
     $ a = \rho_{2D} {\rm ln}(\beta_c \mu_c /\beta \mu) $,
     $  b =  (7 \zeta(3)/8\pi^2) \rho_{2D} \beta^2 $,       
      $  c = (7\zeta(3)/ 8 \pi^2) \rho_{2D} \mu \beta^2  $,
      and $ d =  \rho_{2D} / 4\mu $;  and in the Bose Einstein limit,
                $ a = (1/2) \rho_{2D} {\rm ln}(\mu){\mu_c} $,
     $  b = \rho_{2D} / 8 \mu^2  $,
     $  c = \rho_{2D} / 8 \mid \mu \mid  $ and 
     $  d = \rho_{2D} / 4 \mid \mu \mid  $. 

     Since the values for the GL coefficients are expressible
     analytically only in the two limiting cases we use the full
     expression for these coefficients to calculate the
     susceptibility and take recourse to the numerical evaluation
     of the integrals. 

     Before proceeding further we want to make some remarks regarding
     these coefficients. In the BCS limit the temperature dependence
     of $c$ (which is related to $ \xi^2 $) and the coefficient of the
     fourth order term are similar. (In fact $b$ and $c$ are related
     in both the limits). The 
     reason for this can be traced to the diagrammatic representation
     of these terms. The quadratic term has two internal fermion lines 
     and to get the coefficient $c $ we differentiate $ K^{(1)} $
     twice with respect to momentum. This makes it equivalent to four
     fermion lines which is similar to four fermion lines in the fourth order
     term, particularly when vanishing momentum limit is
     taken. Moreover, $ b $ and $c$ are singular as $ T \sim T_c
     \rightarrow 0 $. This may have interesting consequences when one
     considers the possibility of quantum phase transition in this
     limit. The second remark is regarding the BE limit. In this limit 
     the coefficients, in both two and 
     three dimensions, are independent of temperature. The GL
     expansion look similar to one for system near an
     antiferromagnetic instability \cite{SGM98}. Thus in this limit,
     one can expect that in the limit of T$_c \rightarrow 0$ (in case
     it is possible) a non-Fermi liquid like behavior in some
     transport and thermodynamic properties may be seen and the
     temperature dependence of these quantities will be similar to
     that in antiferromagnets.

     To calculate the pair susceptibility the equation for 
     $\alpha(T, \mu)$ must be solved self-consistently. For this we
     first note that,  
     \begin{eqnarray}
     \frac{1}{\beta} \sum_{q,m}
     \chi(q,\omega_m) &=& \frac{d}{\pi} \int_{-\infty}^\infty
      d\omega \frac{\omega}{(\alpha(T) + c q^2)^2 +
     d^2\omega^2}\frac{1}{(e^{\beta \omega} - 1)} \nonumber \\
     &=& \frac{1}{d \pi} [\ln y - (2y)^{-1} - \psi(y)],
     \end{eqnarray}
     where $\psi(y)$ is the digamma function and $y$ is given by,
     $ (\alpha + c q^2) / 2\pi d T  $.    
     A good interpolation for $\ln y - (2y)^{-1} - \psi(y)$ which is
     valid for large as well as small $y$ is $1/(2y+12y^2)$. Thus, the 
     self-consistent equation for $\alpha(T,\mu)$ can be written as,
   \begin{equation}
   \label{alpha-1}
   \alpha(T, \mu) = a + \frac{b}{d \pi} \int_0^{q_c} dq  \frac{q^{{\rm D}-1}
    }{2y+12y^2} 
   \end{equation}
    where $D$ denotes the dimension of the system. Our interest is
    in how $\alpha(T, \mu)$ which has been obtained within the
    fluctuation theory compares with the mean field value $a$. We
    calculate $\alpha(T, \mu)$ as a function of $\mu$ at temperatures
    just above $T_c$.

    The results of self consistent calculation of Eq. (\ref{alpha-1})
    with the numerical evaluation of the coefficients $b$, $c$
    and $d$ in three and two dimensions is plotted along with the
    corresponding mean field value $a$  in Fig
    (\ref{fig:alpha}). The parameters are taken as $\beta_c = 100$ and
    $\beta = 90$. The behavior of the curves is not very 
    different in two and three dimensions qualitatively. In the
    calculation of the coefficient $a$, we have assumed that $\mu$ at
    temperature just above $T_c$ is the same as its value at $T_c$,
    i.e. $\mu_c$. This is reasonable for temperatures close to
    $T_c$. This is why in the extreme negative $\mu$ regime (the BEC
    regime) $a$ is identically equal to zero 
    since in this regime $a$ is independent of temperature. From the
    figure two things stand out very clearly - the first is that in
    the extreme limits (both the BCS and the BEC), the effect of
    fluctuations is negligible. Secondly, as one moves away from the
    two extreme limits by reducing $ \mid \mu \mid  $, the effect of
    fluctuations increases leading to a dominant contribution to the pair
    susceptibility as  $\mu \rightarrow  0$. In the the BEC regime as
    one moves towards the crossover regime the value of $a$ also
    increases but the fluctuation contribution is much
    larger. It is also apparent that, the fluctuation contribution in
    two dimensions is an order of magnitude higher than in three
    dimensions for the same set of parameters. The curves are shown
    for relatively small transition temperature. Preliminary results
    indicate that as the parameter $ \beta_c $ is reduced from the
    value $ \beta_c = 100 $ towards $ \beta_c = 20 $ corresponding to
    an increase in the transition temperature, the crossover region as
    well as the fluctuation contribution both increase. 

 \begin{figure}
    \begin{center}
    \epsfxsize=8.5 truecm
    \epsfysize=6.0 truecm 
    \centerline{\epsffile{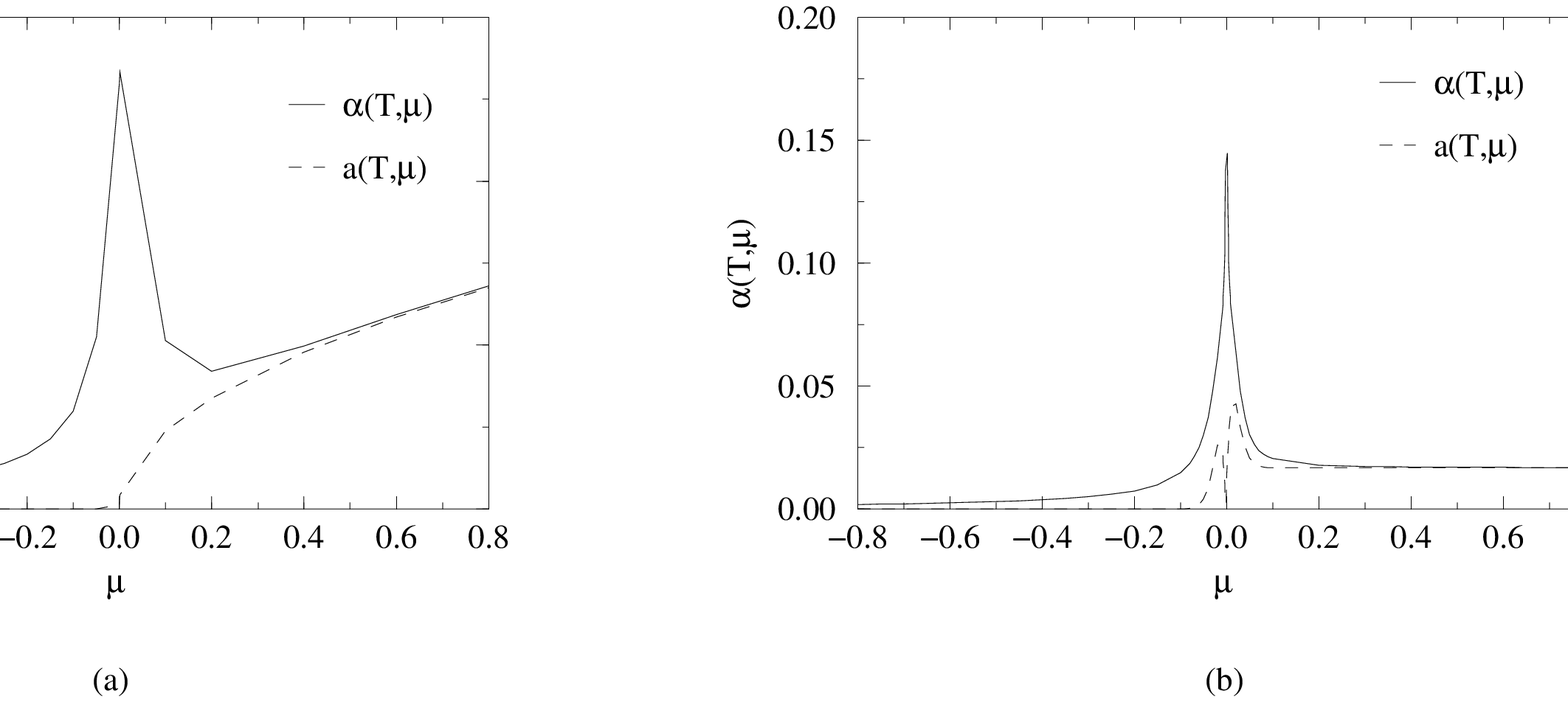}} 

    \vspace{1cm}
    \caption{Inverse susceptibility within the mean fluctuation field 
      approximation ($\alpha(T,\mu)$) and within the mean field approximation 
      ($a(T,\mu)$)plotted as a function of $\mu$ in three dimensions (a) 
       and in two dimensions (b). Note that the scale for the plot in two
       dimensions (b) is one order of magnitude larger than that in three
       dimensions(a).}
    \label{fig:alpha}
    \end{center}
\end{figure}

      In conclusion, we have studied the crossover from
      BCS to BEC from the point of view of pair fluctuations. We find
      that the effects of fluctuations are large in the crossover
      regime, where $\mu$ is close to zero. This shows that a mean
      field calculation will be inadequate in this regime. The
      enhanced fluctuation effects shows that the behavior of physical
      properties can be qualitatively different in this regime. If a
      parallel is drawn with the results of Spin Fluctuation theory,
      then it can be predicted that the physical properties of the
      system may show a non-Fermi liquid behavior if parameters are
      such that one is close to $T_c$ with $T_c \rightarrow 0 $. The
      non-Fermi liquid behavior of some high $T_c$ materials can be
      modeled in this scheme. our discussion has been confined to
      isotropic two- or three- dimensional systems, while the
      anisotropy seem to play an important role in the High $T_c$
      materials. This is especially true for the layered High $T_c$
      compounds. However, in the layered materials, the coherence
      length along the $a-b$ plane is also of the order of that
      expected for the bulk material. Hence, the normal state
      properties of the $a-b$ plane, which are different from the
      properties along the $c$ axis, should in fact follow our two
      dimensional result. The other important point is that the form
      of the pairing potential which we have chosen here is s-wave. It
      is now quite well agreed upon that the pairing potential in most
      of the High $T_c$ materials, especially the layered materials,
      is d-wave. In our opinion the enhancement of fluctuations will
      occur in this case also along with the reflection of the
      inherent anisotropy brought in by the d-wave nature of the pair
      wave function. Work in this direction is in progress.

    \end{document}